\documentclass[aps,preprint]{revtex4}%
\usepackage{amsfonts}
\usepackage{amsmath}
\usepackage{amssymb}
\usepackage{graphicx}%
\providecommand{\U}[1]{\protect\rule{.1in}{.1in}}

\begin{document}
\preprint{ }
\title[Quantum Phase Transition]{Quantum Phase Transition of the Electron-Hole Liquid In the Coupled Quantum
Wells }
\author{V.S. Babichenko}
\affiliation{National Research Center Kurchatov Institute, Moscow, 123182, Russia}
\author{I. Ya. Polishchuk}
\affiliation{National Research Center Kurchatov Institute, Moscow, 123182, Russia}

\begin{abstract}
Many-component electron-hole plasma is considered in the Coupled Quantum Wells
(CQW). It is found that the homogeneous state of the plasma is unstable if the
carrier density is sufficiently small. The instability results in the
breakdown into two coexisting phase - a low-density gas phase and a
high-density electron-hole liquid. The homogeneous state of the electron-hole
liquid is stable if the distance between the quantum wells $\ell$ is
sufficiently small. However, as the distance $\ell$ increases and reaches a
certain critical value $\ell_{cr}$, the plasmon spectrum of the electron-hole
liquid becomes unstable. Hereupon, a quantum phase transition occurs,
resulting in the appearance of the charge density waves of \textit{finite}
amplitude in both quantum wells. The strong mass renormalization and the
strong $Z$-factor renormalization are found for the electron-hole liquid as
the quantum phase transition occurs.

\end{abstract}
\maketitle

\bigskip


\section{Introduction}

For a long time the investigation of the 2D strongly correlated electron
system attracts a great interest of both theorists and experimentalists (see
e.g. Refs. \cite{Grimes, Pudal, Pudal1, Pudal2, Dolgopol, Dolgopol1, Hodel,
Khodel, Hodel1, Volovik, Noz}). The electron hole-plasma (EHP) in the coupled
quantum wells (CQW), where the electrons are localized in one quantum well and
the holes are localized in the other quantum well, occupies a special place
among the low dimensional many-electron systems \cite{23, 24, 25,
Nature-2002-Butov, Nature-2002-Snoke, Larionov2002, Butov2003, Snoke2003,
Butov2004, Dremin2004, Timofeev2005, Levitov2005, Sugakov2006, sugakov2014,
Balatsky, Ando, 28, 29, 30, 34, Rev, sibeldin2016}. The interest in the CQW
has greatly grown in the recent years due to the increasing ability of
manufacturing the high quality quantum well structures in which electrons and
holes are confined in the different spatial regions between which the
tunneling can be made negligible \cite{Rev}. The EHP in the CQW is a
nonequilibrium one, but the electrons and the holes have a large lifetime due
to the spatial separation \cite{23}. Strong electron-hole correlations in such
systems can result in the creation of excitons which are the bound
electron-hole states. A possibility of the exciton Bose-Einstein condensation
(BEC) as well as the superfluidity and the superconductivity in the COW are
considered microscopically in Refs. \cite{23, 24}. The gas-liquid transition,
the features of the liquid exciton phase and the transition into the
superfluid phase are studied as a function of the distance $\ell$ between the
electron and the hole layers in the CQW in Ref.\cite{25}. The strongly
nonideal system of the excitons in the CQW considered as structureless bosons
was considered in \cite{Sugakov2006, Balatsky, sugakov2014}, the exciton
correlation being taken into account in a semi-phenomenological way.

Below we propose a microscopical description of strongly-correlated
\textit{multi-component} electron-hole liquid (EHL) which is a nonideal
multi-component plasma (EHP) in the CQW at zero temperature. The number of
different kinds of the electrons and of the holes is assumed to be large. The
electron-hole system in a many-valley semiconductors is a typical
representative of the multi-component EHP \cite{Ando}. As it was shown for the
first time in Ref. \cite{babich}, the multi-component EHP in bulk
semiconductors possesses the  unconventional Coulomb screening. Such
remarkable feature is connected with occurrence of  characteristic momentum
$p_{0}$ and characteristic energy $\omega_{0}$ which far exceed the
Fermi-momentum $p_{F}$ and the Fermi-energy $\varepsilon_{F},$ respectively.
The parameters $p_{0}$ and $\omega_{0}$ determine the region of the plasmon
spectrum which mainly responsible for the unconventional Coulomb screening in
the multi-component EHP \cite{babich}. Such property of the multi-component
EHP was employed for investigation of various features of the electron-hole
liquid (EHL) \cite{babich1, babich2} .  The features inherent in the
multi-component EHP are also relevant for the multi-component electron gas at
the uniform positive background and the EHP and electronic gas with strong
anisotropic electron spectrum in the quasi-one-dimensional and
quasi-two-dimensional system \cite{babich, keldysh2, Iord1, Iord2,
Pudalov-Ref16, A-Punnoose}.

A possibility of a bulk phase transition of the EHP into EHL was considered in
\cite{keldysh1}, followed by the numerous experimental and the theoretical
investigations (see e.g. \cite{keldysh2, rice}). This phase transition is a
consequence of the instability of the neutral homogeneous EHP if it has a
density smaller than certain critical value $n_{c}.$ The instability results
in the appearance of drops of the EHL with the equilibrium density
$n_{eq}>n_{c}.$ It is remarkable that, if the bulk EHP is a multi-component
one, both $n_{c}$ and $n_{eq}$ are completely determined by the number of the
component $\nu$ \cite{babich}.

The energy of the ground state and the chemical potential of the
multi-component EHP in the CQW were calculated in Refs.\cite{35,36,37} as a
function of the electron density $n$ (which is the same for the holes), the
inter-plane distance $\ell,$ and the number of the components $\nu\gg1$. The
critical concentration $n_{c}=n_{c}\left(  \ell,\nu\right)  $ was found such
that, for the concentration $n<n_{c},$ a homogeneous in-plane charge
distribution is unstable. Such instability resulted in the formation of the
EHL with the equilibrium density $n_{eq}=n_{eq}\left(  \ell,\nu\right)
>n_{c},$ $n_{eq}\sim\nu^{3/2}$ if $\ell\ll\nu^{-1}\ll1,$ and $n_{eq}\sim
\ell^{-3/2}$ if $\ell\gg\nu^{-1}$ \cite{35}. It is shown in Ref. \cite{35}
that, for the density $n=n_{eq},$ the in-plane exciton radius is of the order
of the average distance between the charge carriers within the quantum well.
This fact does not evidence in favour of an existence of exciton as an
structureless particle in the CQW. Instead, strong electron-hole correlations
near the Fermi surface remain. These correlations, in turn, can result in the
unconventional Coulomb screening (inherent in the multi-component EHL), and in
the superconductivity induced due to the Coulomb interaction alone
\cite{babich1, babich2}.

In the present paper we investigate the features of the EHL in the CQW, whose
existence is predicted in \cite{35,36,37}. Like these papers, the system of
units is used in which the effective electron charge $e^{\ast}=e/\sqrt
{\kappa_{0}}$ ($\kappa_{0}$ is the static permittivity), the bare electron
mass $m$ and the Planck constant $\hbar$ are as follows $e^{\ast}=m=\hbar=1.$
For such system of units, the effective Bohr radius, $a_{B}=\hbar^{2}%
/me^{\ast2}=1$ which is taken as a length unit. For the sake of simplicity, we
assume that the masses of electron and hole are equal. As is shown in
\cite{36}, this assumption does not influence the result qualitatively but it
simplifies the calculations significantly. According to Ref. \cite{37}, the
plasmon spectrum of the EHL is stable for $n=n_{eq}$ if $\ell\ll1.$  In this
case, as it is shown in the present paper, both the electron mass and the
$Z$-factor for the Green function experience negligible renormalization
induced by the Coulomb interaction. However, as the distance $\ell$ increases
and reaches a certain critical value $\ell_{cr}$, the plasmon spectrum of the
electron-hole liquid becomes unstable. Hereupon, a quantum phase transition
occurs, resulting in the appearance of the charge density waves of
\textit{finite} amplitude in both quantum wells. The strong mass
renormalization and the strong $Z$-factor renormalization are found for the
electron-hole liquid as the quantum phase transition occurs. All the results
obtained in Refs. \cite{35,36,37} as well as in the present paper are based on
the selection of the diagrams in the small parameters $1/\nu.$ However, the
results obtained seems to be qualitatively valid if the parameter $\nu$ is not
too large. A relationship to the experiments available is considered. 

\section{Green function in the multi-component electron-hole plasma}

The multi-component EHP in the CQW is described with the following Hamiltonian
of the system $\widehat{H}=\widehat{H}_{0}+\widehat{V_{int},}$
\begin{align}
\widehat{H_{0}}  &  =\sum\limits_{\alpha\sigma\mathbf{k}}\frac{k^{2}}%
{2}a_{\alpha\sigma}^{+}\left(  \mathbf{k}\right)  a_{\alpha\sigma}\left(
\mathbf{k}\right)  ,\label{eq.1}\\
~\widehat{V_{int}}  &  =\frac{1}{2S}\sum\limits_{\substack{\alpha
\alpha^{\prime}\sigma\sigma^{\prime}\\\mathbf{kk}^{\prime}\mathbf{q}%
}}V_{\alpha\alpha^{\prime}}\left(  \mathbf{q}\right)  \times a_{\alpha\sigma
}^{+}\left(  \mathbf{k}\right)  a_{\alpha^{\prime}\sigma^{\prime}}^{+}\left(
\mathbf{k}^{\prime}\right)  a_{\alpha^{\prime}\sigma^{\prime}}\left(
\mathbf{k}^{\prime}-\mathbf{q}\right)  a_{\alpha\sigma}\left(  \mathbf{k}%
+\mathbf{q}\right)  .\nonumber
\end{align}
Here $\alpha=e$ stands for the electrons, while $\alpha=h$ stands for the
holes; $\sigma=1,...\nu$ labels the kind of the electron or the hole;
$a_{\alpha\sigma}^{+}\left(  \mathbf{k}\right)  $ and $a_{\alpha\sigma}\left(
\mathbf{k}\right)  $ are the electron or the hole creation and annihilation
operators; $\mathbf{k}$, $\mathbf{k}^{\prime}$, $\mathbf{q}$ are the
$2D-$momenta; $S$ is the area of the QWs. The Coulomb interaction
$V_{\alpha\alpha^{\prime}}$ is assumed to be independent of the kind of the
particle, i.e. of the subscripts $\sigma,$ and
\begin{equation}
V_{\alpha\alpha^{\prime}}\left(  \mathbf{q}\right)  =\left\{
\begin{array}
[c]{c}%
V_{ee}\left(  q\right)  =V_{hh}\left(  q\right)  =V=\frac{2\pi}{q}\text{,
\ \ \ \ \ \ \ }\alpha=\alpha^{\prime}\text{; }\\
V_{eh}\left(  q\right)  =V^{\prime}=-\frac{2\pi}{q}e^{-q\ell}\text{,
\ \ \ }\alpha\neq\alpha^{\prime}%
\end{array}
.\right.  \label{eq.2}%
\end{equation}
A single-particle Green function $G_{\alpha\sigma}\left(  K\right)  $ depends
neither on the subscript $\alpha$ nor on the subscript $\sigma.$ Then,%

\begin{equation}
G_{\alpha\sigma}\left(  K\right)  =G\left(  K\right)  =\left(  i\omega
+\mu-k^{2}/2-\Sigma\left(  K\right)  \right)  ^{-1},\label{3}%
\end{equation}
where $\mu$ is a chemical potential, $\Sigma\left(  K\right)  $ is a
self-energy part (SEP), $K=\left(  i\omega,\mathbf{k}\right)  ,$ $\omega$ is
the Matzubara frequency, $\mathbf{k}$ is the 2D-momentum. Like papers
\cite{35, 36, 37}, the calculation of the Green function is based on the
selection of the diagram in the small parameter $1/\nu\ll1$. Let us represent
the SEP as \ $\Sigma\left(  K\right)  =\Sigma_{H}+\Sigma^{\left(  c\right)
}\left(  K\right)  ,$ where $\Sigma_{H}=2\pi n\ell$ is the $K$-independent
Hartree contribution, and $\Sigma^{\left(  c\right)  }\left(  K\right)  $
involves both the exchange and the correlation contribution. Selecting the
main sequence of the diagram in the parameter $1/\nu$ one obtains for the SEP
$\Sigma^{\left(  c\right)  }\left(  K\right)  $ \cite{35}%
\begin{equation}
\Sigma^{\left(  c\right)  }\left(  P\right)  =-\int\frac{d\omega d^{2}%
k}{\left(  2\pi\right)  ^{3}}U\left(  K\right)  G^{\left(  0\right)  }\left(
\varepsilon+\omega,\mathbf{p}+\mathbf{k}\right)  .\label{eq.4}%
\end{equation}
Here the Green function is $G^{\left(  0\right)  }\left(  K\right)  =\left(
i\omega+p_{F}^{2}/2-k^{2}/2\right)  ^{-1},$ $p_{F}=2\pi^{1/2}\left(
n/\nu\right)  ^{1/2}$ is the Fermi momentum, $\varepsilon_{F}=2\pi n/\nu$ is
the Fermi energy, $n$ is the total concentration (the parameters
$p_{F},\varepsilon_{F},$ $n$ are the same for the electrons and the holes).
The effective interaction reads
\begin{equation}
U\left(  K\right)  =\frac{V\left(  \mathbf{k}\right)  }{1-\chi\left(
\ell\right)  V\left(  \mathbf{k}\right)  \Pi_{0}\left(  K\right)
},\label{eq.4.1}%
\end{equation}
where the polarization operator is given by%

\[
\Pi_{0}\left(  K\right)  =\nu\int\frac{d\omega_{1}d^{2}k_{1}}{\left(
2\pi\right)  ^{3}}G^{\left(  0\right)  }\left(  K+K_{1}\right)  G^{\left(
0\right)  }\left(  K_{1}\right)  .
\]
The function $\chi\left(  \ell\right)  ~$is a monotonic, continuous and slowly
varying one obeying the condition $\chi\left(  \ell\right)  =2$ $~$%
for$~\ell\ll1$ and $\chi\left(  \ell\right)  =1$ for $\ell\gg1$ \cite{35}.

We are interested in the $\Sigma^{\left(  c\right)  }\left(  P\right)  $ for
the momenta and the frequencies which are close to the Fermi ones. On the,
other hand, as shown in \cite{35}, the main contribution into integral
(\ref{eq.4}) originates from the region around the $\omega\sim\omega
_{0}=n^{2/3}/2\gg\varepsilon_{F}$ and $k\sim k_{0}=n^{2/3}\gg p_{F}.$ To
calculate integral (\ref{eq.4}), one should take into account that for $\nu
\gg1$\ the polarization operator $\Pi_{0}\left(  K\right)  $ can be
substituted with its asymptotics for the momentum $k\gg p_{F}$ and the
frequency $\omega\gg\varepsilon_{F}$ as
\begin{equation}
\Pi_{0}\left(  K\right)  =-nk^{2}/\left(  \omega^{2}+\left(  k^{2}/2\right)
^{2}\right)  .\label{largemomentum}%
\end{equation}
Integral (\ref{eq.4}) is readily calculated by substitution
$\mathbf{k\rightarrow}\left(  \chi n\right)  ^{1/3}\mathbf{k}$, $\omega
\rightarrow\left(  \chi n\right)  ^{2/3}\omega.$Then, the calculation of the
$\Sigma^{\left(  c\right)  }\left(  P\right)  $ for $p\ll k_{0},\varepsilon\ll
k_{0}^{2}$ results in the expression for the $\Sigma\left(  P\right)  $ in the
form
\begin{align}
\Sigma\left(  P\right)   &  =\Sigma\left(  0,p_{F}\right)  -i\varepsilon
I_{Z}+\xi_{p}^{\left(  0\right)  }I_{m},\label{eq.6.1}\\
\Sigma\left(  0,p_{F}\right)   &  =2\pi n\ell-C~\left(  \chi n\right)
^{1/3},~\xi_{p}^{\left(  0\right)  }=\left(  p^{2}-p_{F}^{2}\right)  /2
\end{align}
where $I_{Z}=C_{Z}\left(  \chi n\right)  ^{-1/3}$, $I_{m}=C_{m}\left(  \chi
n\right)  ^{-1/3}$, $\xi_{p}^{\left(  0\right)  }=\left(  p^{2}-p_{F}%
^{2}\right)  /2$. The numerical calculation of the constants entering the
$\Sigma\left(  P\right)  $ gives $C\approx1.3,$ $C_{Z}\approx7.6,$ and
$C_{m}=0.4.$

The chemical potential $\mu$ is determined via $\Sigma\left(  0,p_{F}\right)
$ by the well-known relation
\begin{equation}
\mu=p_{F}^{2}/2+\Sigma\left(  0,p_{F}\right)  =2\pi n\left(  1/\nu
+\ell\right)  -C\left(  \chi n\right)  ^{1/3},\label{eq.5}%
\end{equation}
and the Green function reads
\begin{align}
G\left(  P\right)   &  =G\left(  i\varepsilon,p\right)  =\frac{Z}%
{i\varepsilon-\xi_{p}^{\ast}}\text{; \ }\xi_{p}^{\ast}=\frac{p^{2}-p_{F}^{2}%
}{2m^{\ast}};\text{ }\label{eq.7}\\
Z &  =\frac{1}{1+I_{z}}\text{; \ \ \ }\frac{1}{m^{\ast}}=\frac{1+I_{m}%
}{1+I_{z}}<1.\label{eq.7.1}%
\end{align}

For the densities $n<n_{c}=\left[  \frac{C\chi^{1/3}}{6\pi\left(  \ell
+1/\nu\right)  }\right]  ^{3/2},~$one has $\partial\mu/\partial n<0.$ This
fact means an instability of the homogeneous EHP for sufficiently small
densities. Then, chemical potential (\ref{eq.5}) determines the energy per
particle
\begin{equation}
E=\pi\left(  n/\nu\right)  +\pi n\ell-\frac{3}{4}C\left(  \chi n\right)
^{1/3}. \label{eq6}%
\end{equation}
This expression has a minimum for the density
\begin{equation}
n_{eq}=\left[  \frac{C\chi^{1/3}}{4\pi\left(  \ell+1/\nu\right)  }\right]
^{3/2}>n_{c}. \label{eq6-1}%
\end{equation}
The minimum corresponds to the vanishing pressure. For this reason, the
equilibrium state of the EHP at the density $n=n_{eq}$ is the EHL.

Let us consider how the Coulomb interaction affects the effective mass
$m^{\ast}$ of the quasiparticle and the $Z$-factor of the renormalized Green
function for the EHL, i.e. for the density $n=n_{eq}.$ Let $\ell\ll1.$ Then
$n_{eq}\gg1.$ It follows from (\ref{eq.7.1}) that $\Delta m=m^{\ast}-m\ll
m~$and $Z=1-\delta,~\delta\ll1.$ In the opposite case $\ell\gtrsim1,$\ one has
$n_{eq}\sim l^{-3/2}\lesssim1.$ Then, according to (\ref{eq.7.1}), $\Delta
m=m^{\ast}-m\sim m$ and $Z=1-\delta,~\delta\sim1$ and the renormalization is
significant. Thus, the renormalization induced by the Coulomb interaction is
insignificant for $\ell\ll1$ and is visible for $\ell\gtrsim1.$

\section{Vertex Part}

To investigate the plasmon spectrum of the EHL and its stability, let us
calculate the vertex part $\Gamma_{\alpha\sigma,\alpha_{1}\sigma_{1}%
;\alpha^{\prime}\sigma^{\prime},\alpha_{1}^{\prime}\sigma_{1}^{\prime}}$ with
two input fermion ends $\alpha\sigma,\alpha_{1}\sigma_{1}$ and two output
fermion ends $\alpha^{\prime}\sigma^{\prime},\alpha_{1}^{\prime}\sigma
_{1}^{\prime}.$ In what follows, for brevity, we will use the notation
$\Gamma_{\alpha\beta;\alpha^{\prime}\beta^{\prime}}$ instead of $\Gamma
_{\alpha\sigma,\alpha_{1}\sigma_{1};\alpha^{\prime}\sigma^{\prime},\alpha
_{1}^{\prime}\sigma_{1}^{\prime}}.$ Thus, we omit the subscripts
$\sigma,\sigma_{1}\sigma^{\prime},\sigma_{1}^{\prime}$. In particular, the
notation $\alpha,$ in fact, implies $\alpha\sigma.$ This convention reflects
the fact that the value of the vertex part does not depend on the value of the
subscripts $\sigma,\sigma_{1}\sigma^{\prime},\sigma_{1}^{\prime}$ at all.
However, the omitted subscripts should be taken into account when the
summation over such subscripts is necessary.\begin{figure}[ptbh]
\centering\includegraphics[width=0.45\textwidth]{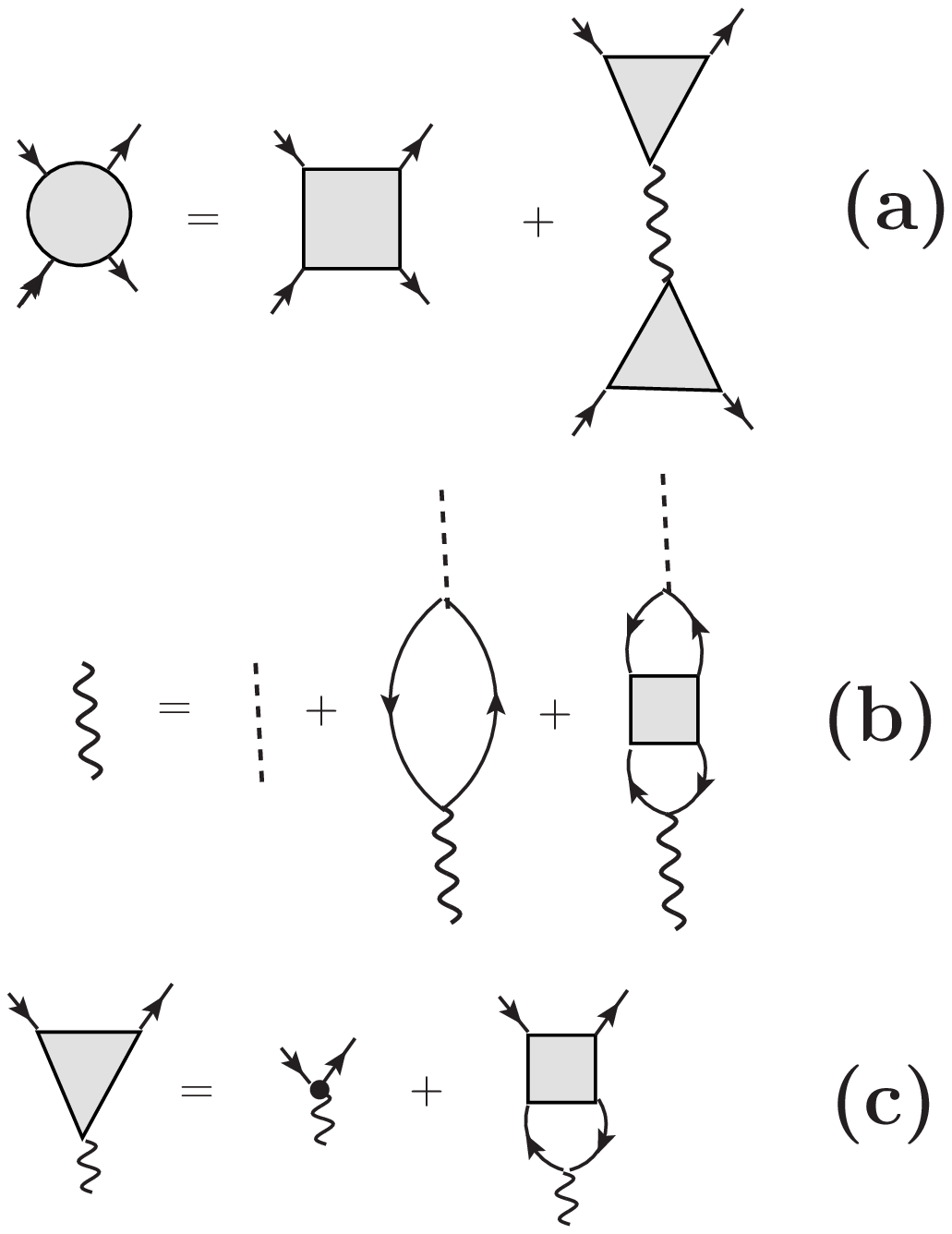}\caption{a) }%
\label{Figure1}%
\end{figure}

The exact diagrammatic representation for the vertex function is given in Fig.
\ref{Figure1}a. In this figure the black circle with two input ends and two
output ends represents the exact vertex part $\Gamma_{\alpha\beta
;\alpha^{\prime}\beta^{\prime}}$; the black square with two input ends and two
output ends represents the \textit{irreducible }vertex part $\overline{\Gamma
}_{\alpha\beta;\alpha^{\prime}\beta^{\prime}}.$ (Any diagram is called an
irreducible one if it can not be cut across one interaction (dotted) line
resulting in two uncoupled parts); the black triangular with one input end,
one output end and one interaction end represents the \textit{irreducible}
vertex part $\overline{\Gamma}_{\alpha\alpha^{\prime};\delta}^{3},$ the wavy
line denotes the effective Coulomb interaction $U_{\delta\delta^{\prime}}.$ In
turn, the effective interaction $U_{\delta\delta^{\prime}}$ is determined by
the self-consistent diagrammatic equation in Fig.\ref{Figure1}b, in which the
dotted lines denote  bare Coulomb interaction (\ref{eq.2}); the inner lines
with arrows denote the exact fermion Green functions. The diagrammatic
equation in Fig.\ref{Figure1}c is an exact relation between the irreducible
vertex parts $\overline{\Gamma}_{\alpha\beta;\alpha^{\prime}\beta^{\prime}}$
and $\overline{\Gamma}_{\alpha\alpha^{\prime};\delta}^{3}$. 

So the analytic representation of the exact diagrammatic equation in Fig.
\ref{Figure1}a is given by
\begin{equation}
\Gamma_{\alpha\beta;\alpha^{\prime}\beta^{\prime}}=\overline{\Gamma}%
_{\alpha\beta;\alpha^{\prime}\beta^{\prime}}+\sum\limits_{\delta_{1}%
;\delta_{2}}\overline{\Gamma}_{\alpha;\alpha^{\prime};\delta_{1}}^{\left(
3\right)  }\cdot U_{\delta_{1};\delta_{2}}\cdot\overline{\Gamma}_{\beta
;\beta^{\prime};\delta_{2}}^{\left(  3\right)  }\label{15}%
\end{equation}
In Fig. \ref{Figure1}b the self-consistent diagrammatic equation reads the
effective interaction $U_{\delta_{1};\delta_{2}},$ which enters Eq.
(\ref{15}). Thus,%

\begin{equation}
U_{\delta_{1};\delta_{2}}\left(  K\right)  =V_{\delta_{1};\delta_{2}}\left(
\mathbf{k}\right)  +\sum\limits_{\rho,\eta}V_{\delta_{1};\rho}\left(
\mathbf{k}\right)  \Pi_{\rho;\eta}\left(  K\right)  U_{\eta;\delta_{2}}\left(
K\right)  . \label{16}%
\end{equation}
Here $\Pi_{\rho;\eta}\left(  K\right)  $ is the exact polarization operator
which, according to Fig. \ref{Figure1}b, reads%

\begin{equation}
\Pi_{\rho;\eta}\left(  K\right)  =\Pi_{0}^{\ast}\left(  K\right)  \delta
_{\rho\eta}+\Pi_{\rho;\eta}^{\left(  c\right)  }\left(  K\right)  . \label{17}%
\end{equation}
In this equation, the polarization operator $\Pi_{0}^{\ast}\left(  K\right)  $
is determined as follows%

\begin{equation}
\Pi_{0}^{\ast}\left(  K\right)  =\nu\int\frac{d\omega_{1}d^{2}k_{1}}{\left(
2\pi\right)  ^{3}}G\left(  K+K_{1}\right)  G\left(  K_{1}\right)  , \label{18}%
\end{equation}
while the $\Pi_{\rho;\eta}^{\left(  c\right)  }\left(  K\right)  $ is
determined via the vertex $\overline{\Gamma}_{\alpha\beta;\alpha^{\prime}%
\beta^{\prime}}$ as is shown in Fig. \ref{Figure1}b. The asymptotic expression
for the $\Pi_{0}^{\ast}\left(  K\right)  $ is%

\begin{align}
\Pi_{0}^{\ast}\left(  K\right)   &  =-\nu\frac{Z^{2}}{2\pi}\frac{\left(
kv_{F}\right)  ^{2}/2}{\omega^{2}+\left(  kv_{F}\right)  ^{2}/2},~k\ll
p_{F}\label{20-1}\\
\Pi_{0}^{\ast}\left(  K\right)   &  =-n\frac{2Z^{2}\xi_{k}^{\ast}}{\omega
^{2}+\left(  \xi_{k}^{\ast}\right)  ^{2}},\text{ }p_{F}\ll k\ll k_{0}\sim
n^{1/3}. \label{20-3}%
\end{align}
\qquad

\begin{figure}[ptbh]
\centering\includegraphics[width=0.45\textwidth]{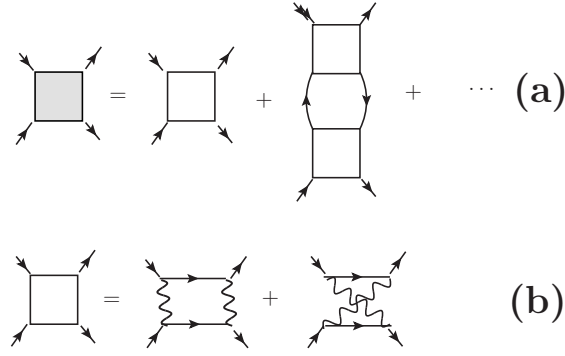}\caption{a) The main
diagrammatic sequence for the $\overline{\Gamma}_{\alpha\beta;\alpha^{\prime
}\beta^{\prime}}$; b) the bare irreducible vertex function $\gamma
_{\alpha\alpha^{\prime}}.$ }%
\label{Figure2}%
\end{figure}

The main diagrammatic sequence for the $\overline{\Gamma}_{\alpha\beta
;\alpha^{\prime}\beta^{\prime}}$ in the parameter $1/\nu\ll1$ is shown in Fig.
\ref{Figure2}a. Let $\gamma_{\alpha\beta;\alpha^{\prime}\beta^{\prime}}$ (the
light square) be irreducible bare vertex part which generates the main
diagrammatic sequence for the vertex part $\overline{\Gamma}_{\alpha
\beta;\alpha^{\prime}\beta^{\prime}}.$ One can show that the $\gamma
_{\alpha\beta;\alpha^{\prime}\beta^{\prime}}$ is composed of two diagrams
shown in Fig. \ref{Figure2}b.

The simple reasoning reveals that $\gamma_{\alpha\beta;\alpha^{\prime}%
\beta^{\prime}}$ may be represented as follows: $\gamma_{\alpha\beta
;\alpha^{\prime}\beta^{\prime}}=\gamma_{\alpha\beta}\delta_{\alpha
\alpha^{\prime}}\delta_{\beta\beta^{\prime}}.$ Then, e.g.,%

\begin{equation}
\gamma_{ee}=\gamma_{hh}=\gamma;~\gamma\left(  p_{1},p_{2},q\right)
=-\int\frac{d^{3}p}{\left(  2\pi\right)  ^{3}}U\left(  p\right)  U\left(
k-p\right)  G\left(  p_{1}-p\right)  \left[  G\left(  p_{2}+p\right)
+G\left(  p_{2}+k-p\right)  \right]  \label{21}%
\end{equation}
The effective interaction $U\left(  K\right)  $ in Eq. (\ref{21}) is given by
Eq. (\ref{eq.4.1}). An analysis of the integrand in Eq. (\ref{21}) (which is
similar to the analysis of the integrand in Eq. (\ref{eq.4})) shows that the
main contribution into integrals (\ref{21}) originates from the region $k\sim
k_{0}\sim n^{1/3}\gg p_{F}$ and $\omega\sim\omega_{0}\sim k_{0}^{2}%
\gg\varepsilon_{F}.$ For this reason, if the components of the external
momenta $p_{1},p_{2},q$ are much smaller than $k_{0},\omega_{0},$ then one can
neglect $p_{1},p_{2},q$ in the integrand. Therefore, the vertex part
$~\gamma\left(  p_{1},p_{2},q\right)  $ does not depend on the $p_{1}%
,p_{2},q.$ After the simple transformation one obtains%
\begin{equation}
\gamma=-\frac{1}{2n^{2}}\int\frac{d^{2}kd\omega}{\left(  2\pi\right)  ^{3}%
}U\left(  K\right)  U\left(  -K\right)  \left(  \Pi_{0}\left(  K\right)
\right)  ^{2}. \label{21-1}%
\end{equation}
To calculate integral (\ref{21-1}), let us take into account that, for $\nu
\gg1,$\ Eq.(\ref{largemomentum}) can be used for the polarization operator
$\Pi_{0}\left(  K\right)  $ for large transfer momentum $k\gg p_{F}.$ Then,
the integral is readily calculated by substitution $\mathbf{k\rightarrow
}\left(  \chi n\right)  ^{1/3}\mathbf{k}$, $\omega\rightarrow\left(  \chi
n\right)  ^{2/3}\omega.$ As a result, one obtains%

\begin{equation}
\gamma=-C_{\gamma}\frac{1}{\left(  \chi n\right)  ^{2/3}},C_{\gamma}\approx0.4
\label{22a}%
\end{equation}

Similarly, for small external momenta one has%

\begin{equation}
\gamma_{eh}=\gamma^{\prime};\gamma^{\prime}=-\frac{1}{2n^{2}}\int\frac
{d^{2}kd\omega}{\left(  2\pi\right)  ^{3}}U^{\prime}\left(  K\right)
U^{\prime}\left(  -K\right)  \left(  \Pi_{0}\left(  K\right)  \right)  ^{2}.
\label{22}%
\end{equation}
Here $U^{\prime}\left(  K\right)  $ is the effective electron-hole
interaction. As is mentioned above, integrals like (\ref{22}) are determined
by the region $k\sim k_{0}\sim n^{1/3}\gg p_{F}$ and $\omega\sim\omega_{0}\sim
k_{0}^{2}\gg\varepsilon_{F}.~$For this region, the integrand is proportional
to $U^{\prime}\left(  K\right)  \sim V\left(  k_{0}\right)  \sim\exp\left(
-k_{0}\ell\right)  .$ In what follows, we are interested in the densities
$n\sim n_{eq}$ (see Eq. (\ref{eq6-1})). In this case for $\ell\ll1$ the
parameter $k_{0}\ell\ll1$ and one has $U^{\prime}\left(  K\right)  =-U\left(
K\right)  .$ In the opposite case $\ell\gg1$ one has $k_{0}\ell\sim\ell
^{1/2}\gg1$ and the integrand in (\ref{22}) vanishes. Thus, we have%
\begin{equation}
\gamma^{\prime}=\gamma\text{ for \ }\ell\ll1\text{; \ \ \ }\gamma^{\prime
}=0\text{ for\ }\ell\gg1\text{\ } \label{22b}%
\end{equation}

Since the bare vertex parts $\gamma$ and $\gamma^{\prime}$ are constant, the
irreducible vertex $\overline{\Gamma}_{\alpha\beta;\alpha^{\prime}%
\beta^{\prime}}$ depends only on the momentum transfer and, thus,
$\overline{\Gamma}_{\alpha\beta;\alpha^{\prime}\beta^{\prime}}=\overline
{\Gamma}_{\alpha\beta;\alpha^{\prime}\beta^{\prime}}\left(  k,\omega\right)
.$The main sequence of the diagram in the parameter $1/\nu$ \ for
$\overline{\Gamma}_{\alpha\beta;\alpha^{\prime}\beta^{\prime}}$ (see Fig.
\ref{Figure2}(a)) is easily sum for $k\ll k_{0}$ and $\omega\ll\omega_{0}.$
Taking into account that the vertex part can be represented in the form
$\overline{\Gamma}_{\alpha\beta;\alpha^{\prime}\beta^{\prime}}=\overline
{\Gamma}_{\alpha\beta}\delta_{\alpha\alpha^{\prime}}\delta_{\beta\beta
^{\prime}},$ where $\overline{\Gamma}_{ee}\left(  K\right)  =\overline{\Gamma
}_{hh}\left(  K\right)  =\overline{\Gamma}\left(  K\right)  $ one has%

\begin{align}
\overline{\Gamma}_{\alpha\beta;\alpha^{\prime}\beta^{\prime}}  &
=\overline{\Gamma}_{\alpha\beta}\left(  K\right)  \delta_{\alpha\alpha
^{\prime}}\delta_{\beta\beta^{\prime}},~K=\left(  i\omega,\mathbf{k}\right)
\label{23}\\
\overline{\Gamma}_{ee}\left(  K\right)   &  =\overline{\Gamma}_{hh}\left(
K\right)  =\overline{\Gamma}\left(  K\right)  =\frac{\gamma-\left(  \gamma
^{2}-\gamma^{\prime2}\right)  \left(  \Pi_{0}^{\ast}\left(  K\right)  \right)
^{2}}{1-2\Pi_{0}^{\ast}\left(  K\right)  \gamma+\left(  \gamma^{2}%
-\gamma^{\prime2}\right)  \left(  \Pi_{0}^{\ast}\left(  K\right)  \right)
^{2}},\\
\overline{\Gamma}_{eh}\left(  K\right)   &  =\overline{\Gamma}^{\prime}\left(
K\right)  =\frac{\gamma^{\prime}}{1-2\Pi_{0}^{\ast}\left(  K\right)
\gamma+\left(  \gamma^{2}-\gamma^{\prime2}\right)  \left(  \Pi_{0}^{\ast
}\left(  K\right)  \right)  ^{2}}.
\end{align}
These expressions are used to calculate the correlation part of the
polarization operator $\Pi_{\rho;\eta}^{\left(  c\right)  }\left(  K\right)  $
(see Eq. (\ref{17})) and the vertex $\overline{\Gamma}_{\alpha,\alpha^{\prime
};\delta}^{\left(  3\right)  }$ (see. Fig. \ref{Figure1}c). As a result, one obtains%

\begin{equation}
\Pi_{\rho\eta}^{\left(  c\right)  }\left(  K\right)  =\Pi_{0}^{\ast}\left(
K\right)  \overline{\Gamma}_{\rho\eta}\Pi_{0}^{\ast}\left(  K\right)  ,
\label{25}%
\end{equation}

\begin{equation}
\overline{\Gamma}_{\alpha,\alpha^{\prime};\delta}^{\left(  3\right)
}=\overline{\Gamma}_{\alpha;\delta}^{\left(  3\right)  }\delta_{\alpha
\alpha^{\prime}},\overline{\Gamma}_{\alpha;\delta}^{\left(  3\right)  }\left(
K\right)  =\delta_{\alpha\delta}+\overline{\Gamma}_{\alpha;\delta}\left(
K\right)  \Pi_{0}^{\ast}\left(  K\right)  .\label{26}%
\end{equation}
Substituting (\ref{23}), (\ref{25}), (\ref{26}) into Eq. (\ref{15}) for the
vertex $\Gamma_{\alpha\beta;\alpha^{\prime}\beta^{\prime}},$ one has
$\Gamma_{\alpha\beta;\alpha^{\prime}\beta^{\prime}}=\Gamma_{\alpha\beta}%
\delta_{\alpha;\alpha^{\prime}}\delta_{\beta;\beta^{\prime}},$ where%

\begin{equation}
\Gamma_{ee}=\Gamma_{hh}=\Gamma\left(  K\right)  =\frac{\left(  V+\gamma
\right)  -\left[  \left(  V+\gamma\right)  ^{2}-\left(  V^{\prime}%
+\gamma^{\prime}\right)  ^{2}\right]  \Pi_{0}^{\ast}}{\left[  1-\left(
V+V^{\prime}+\gamma+\gamma^{\prime}\right)  \Pi_{0}^{\ast}\right]  \left[
1-\left(  V-V^{\prime}+\gamma-\gamma^{\prime}\right)  \Pi_{0}^{\ast}\right]
}, \label{27}%
\end{equation}

\begin{equation}
\Gamma_{eh}=\Gamma^{\prime}\left(  K\right)  =\frac{V^{\prime}+\gamma^{\prime
}}{\left[  1-\left(  V+V^{\prime}+\gamma+\gamma^{\prime}\right)  \Pi_{0}%
^{\ast}\right]  \left[  1-\left(  V-V^{\prime}+\gamma-\gamma^{\prime}\right)
\Pi_{0}^{\ast}\right]  }. \label{28}%
\end{equation}

\section{Plasmon spectrum and instability}

Let us investigate the plasmon spectrum of the EHP in the CQW which is
determined by poles of vertex parts (\ref{27}) and (\ref{28}). First, let us
consider the case $l\ll1$. Then, it follows from (\ref{27}), (\ref{28}) that%

\begin{equation}
\Gamma=\frac{\left(  V-\pi\ell\right)  }{\left[  1-2\left(  V-\pi\ell\right)
\Pi_{0}^{\ast}\right]  }+\frac{\left(  \gamma+\pi\ell\right)  }{\left[
1-2\left(  \gamma+\pi\ell\right)  \Pi_{0}^{\ast}\right]  }, \label{29}%
\end{equation}

\begin{equation}
\Gamma^{\prime}=\frac{-\left(  V-\pi\ell\right)  }{\left[  1-2\left(
V-\pi\ell\right)  \Pi_{0}^{\ast}\right]  }+\frac{\left(  \gamma+\pi
\ell\right)  }{\left[  1-2\left(  \gamma+\pi\ell\right)  \Pi_{0}^{\ast
}\right]  }.\label{30}%
\end{equation}
Let us substitute (\ref{20-1}) into (\ref{29}) and (\ref{30}) and replace the
Matzubara frequency $i\omega$ by the real frequency $\omega.$ The pole of the
vertex parts $\Gamma$ and $\Gamma^{\prime}$ is given by the second terms in
Eq. (\ref{29}) or Eq.(\ref{30}). Then, the plasmon spectrum is determined by
equation
\begin{equation}
1+2\left(  \gamma+\pi\ell\right)  \frac{\nu}{2\pi}\frac{\left(  kv_{F}\right)
^{2}/2}{-\omega^{2}+\left(  kv_{F}\right)  ^{2}/2}=0.\label{31}%
\end{equation}
The spectrum is stable if $\omega,$ which obey Eq. (\ref{31}), is real. This
takes place if%

\begin{equation}
n>\left[  \frac{C_{\gamma}}{\chi^{2/3}\left(  \frac{1}{\nu}+l\right)
}\right]  ^{3/2}.\label{32}%
\end{equation}
Thus, if $n>n_{cr}=\left[  \frac{C\chi^{1/3}}{6\pi\left(  1/\nu+\ell\right)
}\right]  ^{3/2}$, the plasmon spectrum is stable. In the opposite case,
$n<n_{cr},$ the pole takes place for imaginary $\omega$. This denotes an
instability of the plasmon spectrum. This instability just corresponds to the
thermodynamic instability of the homogeneous EHP for the densities $n<n_{cr}$
for which one has $\partial\mu/\partial n<0$ \cite{babich}. Let us pay
attention that for $\ell\ll1,$ the plasmon spectrum remains stable for the
equilibrium EHL which has the density $n_{eq}=\left[  \frac{C\varkappa^{1/3}%
}{4\pi\left(  1/\nu+\ell\right)  }\right]  ^{3/2}>n_{c}.$\ 

Now let us investigate the plasmon spectrum for the case $\ell\gg1.$ Let us
consider momenta and frequencies which obey the limitations $\left(
1/\ell\right)  \ll k\ll k_{0}$ , $\omega\ll\omega_{0}$. In this case
$V^{\prime}\left(  k\right)  $ vanishes. Also, according to (\ref{22b}),
$\gamma^{\prime}=0.$ Therefore, it follows from (\ref{28}) that $\Gamma
^{\prime}=0.$ So in the case $\ell\gg1$ Eq. (\ref{27}) reads%

\begin{equation}
\Gamma\left(  K\right)  =\frac{V\left(  k\right)  +\gamma}{1-\Pi_{0}^{\ast
}\left(  K\right)  \left(  V\left(  k\right)  +\gamma\right)  }. \label{33}%
\end{equation}
Let us substitute $V,\Pi_{0}^{\ast},$ $\gamma$ by expressions Eqs.
(\ref{eq.2}), (\ref{20-3}), (\ref{22a}) for the momentums $p_{F}\ll k\ll
k_{0}$ and change $i\omega$ by $\omega$ As a result, one obtains%

\begin{equation}
\Gamma\left(  K\right)  =\frac{\left(  \frac{2\pi}{k}-\frac{C_{\gamma}%
}{n^{2/3}}\right)  \left(  \omega^{2}-\left(  \xi_{k}^{\ast}\right)
^{2}\right)  }{\omega^{2}-\xi_{k}^{\ast}\left(  \xi_{k}^{\ast}+nZ^{2}\left(
\frac{2\pi}{k}-\frac{C_{\gamma}}{n^{2/3}}\right)  \right)  }.\label{34}%
\end{equation}
A pole of the $\Gamma\left(  K\right)  $ determines the plasmon spectrum and
exists for the frequencies%
\begin{equation}
\omega_{p}^{2}\left(  k\right)  =\xi_{k}^{\ast}\left(  \xi_{k}^{\ast}%
+nZ^{2}\left(  \frac{2\pi}{k}-\frac{C_{\gamma}}{n^{2/3}}\right)  \right)
.\label{35}%
\end{equation}
Let us investigate a behavior of the plasmon spectrum for the EHL of a density
$n\sim n_{eq}.$ One can easy see that, for small momentums $k\ll k_{0}$ which
additionally belong to the interval $n^{2/3}\lesssim k\lesssim n^{1/2},$ the
plasmon frequency $\omega_{p}\left(  k\right)  $ becomes imaginary. This means
an instability of the homogeneous state of the  EHL with respect to an
appearance of the spatially inhomogeneous periodic in-plane charge
distribution with a period characterized by the wave vector $k.$ Such a charge
density fluctuation  describes the charge density waves (CDW), which are
in-phase for the electron and the hole layers. For the equilibrium EHL with
$n\sim n_{eq}\sim\ell^{-3/2},$ one has $\frac{1}{\ell}\lesssim k\lesssim
\frac{1}{\ell^{3/4}}$ and, thus, the period of the CDW obeys the condition
$\ell^{3/4}\lesssim D\lesssim\ell.$

\section{Conclusion}

Thus, for $\ell\ll1$ the homogeneous state of the EHL with the density
$n_{eq}\sim\nu^{3/2}$ is stable. However, as the distance $\ell$\ increases,
the plasmon spectrum becomes softer for finite momenta $k\gtrsim\frac{1}{\ell
}$. Then, for a certain $\ell_{cr}\sim1,$ there appears a momentum
$k=k_{cr}=1/\ell_{cr}$ for which the plasmon frequency vanishes. As the
distance $\ell$ increases, the plasmon frequencies characterized by the
wave-vector interval $\frac{1}{\ell}\lesssim k\lesssim\frac{1}{\ell^{3/4}}$
become imaginary. As a result, the CDW appears. This feature of the plasmon
spectrum implies that EHL in the CQW experiences a quantum phase transition in
the parameter $\ell.$

Note that some of the results obtained above are valid for multi-component
electron gas at the positive background \cite{Grimes, Pudal, Pudal1,
Pudal2,Dolgopol1, Iord1, Iord2, Pudalov-Ref16, A-Punnoose}. In particular,
this concerns the effective mass renormalization, the $Z$-factor
renormalization for the single-particle Green function, the dependence of the
ground state energy and the chemical potential of the electrons. Also, the
conclusion remains valid and leads to an instability of the ground state of
the electron gas with respect to appearance of the CDW for sufficiently small
density. However, a significant difference takes place: in contrast to EHL the
electron gas at the positive background cannot find the equilibrium density to
minimize the ground state energy since the electron density is settled by the
positive background.

There are several experiments in which the EHL is seemed to be observed in CQW
\cite{sibeldin44, sibeldin45, sibeldin46}. The result obtained in the present
paper may be verified for the systems investigated in these papers.

\vfill\eject

\end{document}